\documentclass[3p,preprint,number]{elsarticle}
\DeclareGraphicsExtensions{.pdf,.gif,.jpg,.pgf}

\usepackage[english]{babel}
\usepackage{ragged2e}
\usepackage{blindtext}
\usepackage{mathtools}
\usepackage{lipsum}
\usepackage{array}
\usepackage{colortbl}
\usepackage{chemfig}
\usepackage{tikz}
\usepackage{pgfplots}
\usepackage{amsmath}
\usepackage{physics}
\pgfplotsset{compat=1.7}
\usepackage{amssymb}
\usepackage{subcaption}
\usepackage{siunitx}
\usepackage{enumitem}
\usepackage{xcolor}
\usepackage{multirow}
\usepackage{adjustbox}
\usepackage[latin1]{inputenc}
\usepackage{csquotes}
\usetikzlibrary{patterns}
\usepackage[english]{babel}

\makeatletter
\newcommand*{\rom}[1]{\expandafter\@slowromancap\romannumeral #1@}
\makeatother
\usepackage{tikz}
\usetikzlibrary{matrix}
\graphicspath{ {c:/Users/User/Documents/PhD/Year 1/LaTeX/Figures/} }
\usepackage{blindtext}
\usepackage{titlesec}
\titleformat{\chapter}[display]   
{\normalfont\huge\bfseries}{\chaptertitlename\ \thechapter}{16pt}{\Huge}   
\titlespacing*{\chapter}{0pt}{-10pt}{5pt}
\usepackage[longend,ruled]{algorithm2e} 
\usepackage{layouts}
\usepackage{changes} 

\definechangesauthor[name=James Mckenna, color=blue]{JM}

\begin{document}
	\bibliographystyle{elsarticle-num.bst}
	{\setlength{\parindent}{0cm} 
	\begin{frontmatter}
		\title{A New Riemann Solver for Modelling Bridges in Flood Flows - Development and Experimental Validation}
		\author{James Mckenna}
		\ead{J.Mckenna4@newcastle.ac.uk}
		\author{Vassilis Glenis\corref{cor1}}
		\ead{Vassilis.Glenis@newcastle.ac.uk}
		\author{Chris Kilsby}
		\ead{Chris.Kilsby@newcastle.ac.uk}
	
		\cortext[cor1]{Corresponding author}
		\address{School of Engineering, Newcastle University, Newcastle upon Tyne, United Kingdom}
		\date{Submitted 29th June 2022}
		\begin{abstract}
			Flows in rivers can be strongly affected by obstacles to flow or artificial structures such as bridges, weirs and dams. This is especially true during floods, where significant backwater effects or diversion of flow out of bank can result. However, within contemporary industry practice, linear features such as bridges are often modelled using coarse approximations, empirically based methods or are omitted entirely. Presented within this paper is a novel Riemann solver which is capable of modelling the influence of such features within hydrodynamic flood models using finite volume schemes to solve the shallow water equations. The solution procedure represents structures at the interface between neighbouring cells and uses a combination of internal boundary conditions and a different form of the conservation laws in the adjacent cells to resolve numerical fluxes across the interface. Since the procedure only applies to the cells adjacent to the interface at which a structure is being modelled, the method is therefore potentially compatible with existing hydrodynamic models. Comparisons with validation data collected from a state of the art research flume demonstrate that the solver is suitable for modelling a range of flow conditions and structure configurations such as bridges and gates. 
		\end{abstract}
	\end{frontmatter}
	\section{Introduction}
		Hydrodynamic models are a vital component of contemporary flood risk management practice, providing evidence to inform future investment and risk assessment. It follows that the accurate modelling of hydrodynamic phenomena is fundamental to effective flood risk management since the quantification and analysis of risk is dependent on modelling results. Channel structures, such as bridges, weirs and gates can act as partial barriers to flow, significantly influencing the local flow characteristics \cite{RN93,RN92,RN90}. Yet, despite the established importance of accurately capturing hydrodynamic phenomena, methods for modelling linear features which act as partial barriers in two-dimensional computational domains are relatively under-developed. Methods for modelling such interactions within three-dimensional simulations exist \cite{RN263}, however, the complex meshing requirements and the inherent computational expense prohibits their use on spatial domains relevant to flood risk management. Moreover, within contemporary industry practice such features are often modelled using coarse approximations, empirically based methods or by omitting such features entirely \cite{RN264,RN265,RN272,RN273}. There is no unified approach to modelling linear features within horizontal two-dimensional hydrodynamic models, with available methods including:
		\begin{itemize}[noitemsep]
			\item Modelling as finite bottom elevations.
			\item Local friction-based representation.
			\item Modelling as `holes' within the mesh.
			\item Modelling via internal boundary conditions.
			\item Modelling as source terms within the numerical scheme.
			\item Modification of the conservation laws.
		\end{itemize}
		However, each method has its respective limitations. A study conducted by Alcrudo \cite{RN266} demonstrated significant challenges related to modelling obstructions with local friction-based representations. This was primarily due to the impracticality of selecting an appropriate Manning\textquotesingle s roughness coefficient. The use of finite bottom elevations was also found to be problematic due to numerical instabilities for some numerical schemes. Both methods are coarse, unphysical approximations, making them unsuitable for high resolution modelling of hydrodynamic interactions with hydraulic structures. Modelling features as \lq holes \rq within the mesh, commonly referred to as the mesh discretisation method, is also unsuitable for generalised treatment of linear features since the method is incompatible with partial barriers to flow. This is due to the fact that once a feature is modelled as a hole within the mesh, the feature is no longer part of the computational domain, disrupting the exchange of conserved variables (depth and momentum for the shallow water equations) across the feature.
		
		More sophisticated approaches include the work of Maranzoni et al. \cite{RN261}, who locally modified the conservation laws by introducing a fictitious vertical slot to the ceiling of the cells representing a linear feature. This enables the transition between free surface and pressurised flow to be captured via the Preissmann slot concept \cite{RN267}. However, this also means that the method suffers from the well documented limitations of the Preissmann slot concept as discussed by Vasconcelos, Wright and Roe \cite{RN248} and Malekpour and Karney \cite{RN262} , which the authors also acknowledge in their work. Furthermore, the method requires further work in order to be suitable for generalised treatment of linear features, as it can only currently simulate pressurisation without overtopping.
		
		Morales-Hern\`andez et al. \cite{RN107}, Ratia et al. \cite{RN102} and Dazzi et al. \cite{RN101} have all proposed implementations of internal boundary conditions to model hydraulic structures. The proposed methods calculate the unit discharge across hydraulic structures by selecting appropriate rating curves or discharge formulae dependent on the inundation depth calculated from the adjacent cells or at the interface. Morales-Hern\`andez et al. \cite{RN107} and Dazzi et al. \cite{RN101} both present similar methods with the PARFLOOD model presented by Dazzi et al. promising the most general applicability. However, the reliance on stage-discharge relationships under certain circumstances, sensitivity to the selection of uncertain discharge coefficients and the potential to induce directionality for skew features are conceivable flaws. Furthermore, as shown by Hou and Le Floch \cite{RN268}, use of non-conservative schemes can result in convergence to incorrect solutions.
		
		The head loss source term introduced by Ratia et al. \cite{RN102} accounts for the losses induced by contraction and expansion of flow through a hydraulic structure. Whilst this method is effective in capturing the macroscale effects induced by the presence of the structure, it is unable to sufficiently capture near-field flow effects including potential downstream supercritical flows. It is therefore clear that more progress is required towards the development of a solver capable of accurately resolving fluxes across general partial barriers to flow, within two-dimensional finite volume schemes, for the purpose of flood modelling. Consequently, the purpose of this paper is to build upon some of these methods, presenting a novel method capable of generalised treatment of partial barriers to flow. The method is intended to be compatible with existing flood models utilising Finite Volume (FV) schemes to solve the shallow water equations. 
		\section{Mathematical Model}
		The proposed method utilises internal boundary conditions in addition to using a different form of the conservation laws in the adjacent cells. As a result, hydraulic structures are idealised as existing at the interface and modelled as a partially reflective boundary between the adjacent cells. Consequently the interfaces within the computational domain can be divided into structure interfaces, employing the novel solution procedure to resolve numerical fluxes, and non-structure interfaces, employing a standard solution procedure to resolve numerical fluxes.
		
		For the non-structure interfaces and the corresponding adjacent cells, a one-dimensional (1D) FV scheme is used to solve the 1D Shallow Water Equations (1D-SWE) given as:
		\begin{equation}
			\textbf{U}_t + \textbf{F}(\textbf{U})_x=\textbf{S}(\textbf{U})
		\end{equation}
		Where $\textbf{U}$ is the vector of conserved variables, $\textbf{F}(\textbf{U})$ is the vector of fluxes and $\textbf{S}(\textbf{U})$ is a vector of sources comprising of $\textbf{S}_0$, the bed slope source term and $\textbf{S}_f$, the bed friction source term. These terms are given as follows:
		\begin{gather}
			\textbf{U} = \begin{bmatrix}
				h \\[6pt]
				hu \\
			\end{bmatrix} \textrm{ , }
			\textbf{F} = \begin{bmatrix}
				hu \\[6pt]
				hu^2 + \frac{1}{2}gh^2 \\
			\end{bmatrix} \textrm{ , }
			\textbf{S}_0 = \begin{bmatrix}
				0 \\[6pt]
				-gh\frac{\partial z}{\partial x} \\
			\end{bmatrix} \textrm{ , }
			\textbf{S}_f = \begin{bmatrix}
				0 \\[6pt]
				- \tau_f \\
			\end{bmatrix}
		\end{gather}
		Whereby $h$ denotes the depth of flow, $z$ is the elevation of the bed, $u$ denotes the velocity component in the $x$ direction, $g$ is the acceleration due to gravity and $\tau_f$ is the shear stress due to friction in accordance with Manning's equation: $$\tau_f = C_fu|u| = \frac{gn^2}{\sqrt[3]{h}}u|u|$$ where $n$ is Manning's roughness coefficient.
		For the structure interfaces and corresponding adjacent cells, a 1D FV scheme is used to solve the two layer 1D-SWE as derived by Spinewine et al. \cite{RN39}:
		\begin{equation}\label{eq: 2-layer SWE 1}
			\textbf{U}_t + \textbf{F}(\textbf{U})_x=\textbf{S}(\textbf{U})
		\end{equation}
		\begin{gather}
			\textbf{U} = \begin{bmatrix}
				h_2 \\
				h_2u_2 \\
				h_1 \\
				h_1u_1 
			\end{bmatrix} \\
			\textbf{F} = \begin{bmatrix} \label{eq: 2-layer SWE 2}
				h_2u_2 \\
				\frac{(h_2u_2)^2}{h_2} + \frac{1}{2}gh_2^2 \\
				h_1u_1 \\
				\frac{(h_1u_1)^2}{h_1} + \frac{1}{2}gh_1^2 + \chi gh_2h_1 \\
			\end{bmatrix}
			= \begin{bmatrix}
				q_2 \\
				\sigma_2 \\
				q_1 \\
				\sigma_1 
			\end{bmatrix} \\
			\textbf{S} = \begin{bmatrix}
				0 \\
				\frac{R_{12}}{\rho_2} \\
				0 \\
				-\frac{\chi R_{12}}{\rho_2}+R_1 - \tau_f
			\end{bmatrix}
			= \begin{bmatrix}
				0 \\
				-gh_2\frac{\partial z_1}{\partial x} \\
				0 \\
				\chi g h_2 \frac{\partial z_1}{\partial x} - g(\chi h_2 + h_1)\frac{\partial z_0}{\partial x} - \tau_f
			\end{bmatrix}
		\end{gather}
		Where the subscript $2$ refers to the upper layer and subscript $1$ refers to the lower layer of a two layer shallow water model. $R_{12}$ is the reaction force exerted by the lower layer onto the upper layer, $R_{1}$ is the reaction of the bed onto the bottom layer and $\rho_k$ is the density of the fluid in layer $k$. 
		Both sets of equations are discretised using the same first order accurate, explicit FV scheme whereby the conserved variables are updated in accordance with equation (\ref{eq: Conservative Update}).
		\begin{equation}\label{eq: Conservative Update}
			\textbf{U}^{n+1}_i = \textbf{U}^{n}_i - \frac{\Delta t}{\Delta x}\left[\textbf{F}_{i+\frac{1}{2}} -  \textbf{F}_{i-\frac{1}{2}}\right] + \Delta t \textbf{S}\left(\textbf{U}_i^n \right)
		\end{equation}
		Where the subscript $i$ represents the $i$th cell, the superscript $n$ represents the $n$th time level and $\Delta x$ and $\Delta t$ represent the cell size and time step respectively. Although a 1D scheme is implemented in this case, implementation as a 2D scheme requires no fundamental changes to the method. 
		\subsection{Numerical Flux Computation}
		Harten-Lax-van Leer (HLL) approximate Riemann solvers \cite{RN269} are used to resolve intercell numerical fluxes for the entire computational domain, however, other approximate Riemann solvers may be used to resolve numerical fluxes across non-structure interfaces. A novel solution procedure is proposed for resolving numerical fluxes across structure interfaces using a newly developed HLL approximate Riemann solver.
		
		The fundamental concept behind the proposed method is the assumption that the motion of the fluid can be treated as primarily horizontal in nature, which is a necessary condition required for application of the shallow water equations. Consequently, the structure cells can be divided into horizontal layers corresponding to the structure idealised at the interface. By dividing the layers in such a way, component fluxes can be resolved for each layer. Through summation of the component fluxes at an interface, it is then possible to determine a flux for the left and right sides of the structure interface. The partially reflective nature of the boundary is captured by implementing appropriate boundary conditions for the individual layers as well as discretising the cell average properties of the respective left and right states (see Figure \ref{fig: Structure Cells}). Based on the boundary condition applied to the layer at the interface the layers are designated as either `\textit{open}', corresponding to a transmissive boundary condition at the interface, or `\textit{closed}', corresponding to a reflective boundary condition at the interface. Structure interfaces require a left and right flux, as opposed to a single flux for non-structure interfaces, due to the fact that the left and right states for closed layers are considered to be discontinuous. 
		Fluxes for submerged layers are calculated using (\ref{eq: Submerged Flux}).
		\begin{gather}\label{eq: Submerged Flux}
			\textbf{F} = \begin{bmatrix}
				h_ku_k \\
				\frac{(h_ku_k)^2}{h_k} + \frac{1}{2}gh_k^2 + \chi gh_{u}h_k \\
			\end{bmatrix}
			= \begin{bmatrix}
				q_k \\
				\sigma_k 
			\end{bmatrix}
		\end{gather}
		Where the subscript $k$ represents the $k$th layer and $h_u$ represents the sum of the layer depths above the submerged layer. Since the layers are the same density the $\chi=\frac{\rho_1}{\rho_2}$ term is always equal to one. Fluxes for free-surface layers (the upper most layer of a structure cell) are calculated using (\ref{eq: Free Surface Flux}).
		\begin{gather}\label{eq: Free Surface Flux}
			\textbf{F} = \begin{bmatrix}
				h_ku_k \\
				\frac{(h_ku_k)^2}{h_k} + \frac{1}{2}gh_k^2 \\
			\end{bmatrix}
			= \begin{bmatrix}
				q_k \\
				\sigma_k 
			\end{bmatrix}
		\end{gather}
		\begin{figure}[hbt!]
			\centering
			\begin{tikzpicture}
				\draw[thick, ->] (-0.75,0) -- (-0.75,2.5) node[above] {$h$};
				
				\draw[thick] (0,0) -- (12,0); 
				\fill[fill=blue!25!white, draw=black] (0,0) rectangle (2,2);
				\draw[dashed] (2,0) -- (2,2.5) node[above] {NSI};
				\draw[thick] (0,-0.5) -- (0,2.5) node[above] {$a$};
				\fill[fill=blue!25!white, draw=black] (2,0) rectangle (4,2);
				\draw[dashed] (4,0) -- (4,2.5) node[above] {NSI};
				\draw[dashed] (6,0) -- (6,2.5) node[above] {SI};
				
				\fill[fill=blue!60!white, draw=black] (4,0) rectangle (6,0.5); 
				\node[below] at (5,0.5) {\large Open};
				\fill[fill=blue!60!white, draw=black] (6,0) rectangle (8,0.5); 
				\node[below] at (7,0.5) {\large Open};
				\fill[fill=blue!40!white, draw=black] (4,0.5) rectangle (6,1.5); 
				\node[below] at (5,1.25) {\large Closed};
				\fill[fill=blue!40!white, draw=black] (6,0.5) rectangle (8,1.5); 
				\fill[fill=black!25!white, draw=black] (5.9,0.5) rectangle (6.1,1.5); 
				\node[below] at (7,1.25) {\large Closed};
				\fill[fill=blue!20!white, draw=black] (4,1.5) rectangle (6,2); 
				\node[below] at (5,2) {\large Open};
				\fill[fill=blue!20!white, draw=black] (6,1.5) rectangle (8,2); 
				\node[below] at (7,2) {\large Open};
				
				\draw[dashed] (8,0) -- (8,2.5) node[above] {NSI};
				\fill[fill=blue!25!white, draw=black] (8,0) rectangle (10,2);
				\draw[thick] (12,-0.5) -- (12,2.5) node[above] {$b$};
				\draw[dashed] (10,0) -- (10,2.5) node[above] {NSI};
				\fill[fill=blue!25!white, draw=black] (10,0) rectangle (12,2);
				
				\draw [decorate, decoration={brace,amplitude=6pt,raise=0pt}, thick] (8,0) -- (4,0); 
				\node[below] at (6,-0.15) {\large Structure Cells};
				\draw [decorate, decoration={brace,amplitude=6pt,raise=0pt}, thick] (4,0) -- (0,0); 
				\node[below] at (2,-0.15) {\large Non-Structure Cells};
				\draw [decorate, decoration={brace,amplitude=6pt,raise=0pt}, thick] (12,0) -- (8,0); 
				\node[below] at (10,-0.15) {\large Non-Structure Cells};
			\end{tikzpicture}
			\caption{A simple computational domain $[a,b]$ illustrating the designation of structure and non-structure cells as well as open and closed layers for the structure cells. Non-structure interfaces are denoted using the abbreviation NSI and structure interfaces are denoted using the abbreviation SI. The split layer properties of the structure cells are only required for computing intercell fluxes across structure interfaces; a non-structure interface that is adjacent to a structure cells treats the structure cell as a non-structure cell utilising the cell-average properties.}
			\label{fig: Structure Cells}
		\end{figure}
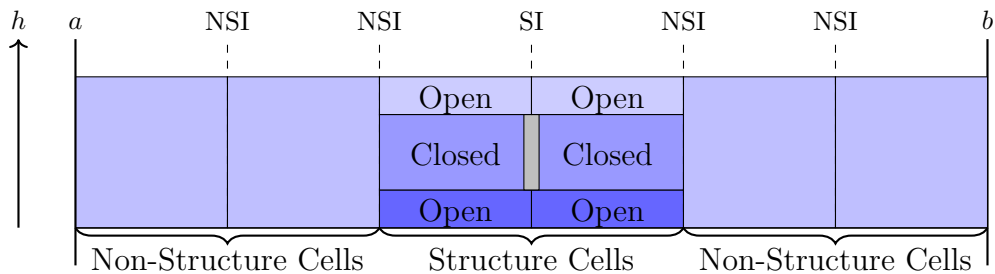
		\subsection{Solution Procedure}
		At each timestep the adjacent cells are divided into horizontal layers corresponding to the base and cover of the idealised structure at the interface, as demonstrated in Figure \ref{fig:Gate Cases}. This corresponds with the prior outlined assumption that the direction of flow is primarily parallel to the bed. Whilst there is undoubtedly a vertical exchange in momentum between the horizontal layers at such an interface, it is impractical to model such effects within a one-dimensional or two-dimensional scheme.
		\begin{figure}[hbt!]
			\centering
			\begin{tikzpicture}
				\draw[thick, ->] (-0.75,8) -- (-0.75,11) node[above] {$h$};
				\node[above] at (1.5,11) {\textbf{Case \rom{1}}};
				\fill[fill=blue!60!white, draw=black] (0,8) rectangle (1.5,8.75); 
				\fill[fill=blue!60!white, draw=black] (1.5,8) rectangle (3,8.5); 
				\draw[thick, ->] (0,8) -- (3.25,8) node[right] {$x$};
				\draw[thick] (1.5,8) -- (1.5,10.5);
				\fill[fill=black!25!white, draw=black] (1.4,9) rectangle (1.6,10.5); 
				\draw[dashed] (-0.25,9) -- (3.25,9) node[right] {$z_1$};
				\node[above] at (6,11) {\textbf{Case \rom{2}}};
				\fill[fill=blue!60!white, draw=black] (4.5,8) rectangle (6,9); 
				\fill[fill=blue!40!white, draw=black] (4.5,9) rectangle (6,10); 
				\fill[fill=blue!60!white, draw=black] (6,8) rectangle (7.5,8.75); 
				\draw[thick, ->] (4.5,8) -- (7.75,8) node[right] {$x$};
				\draw[thick] (6,8) -- (6,10.5);
				\fill[fill=black!25!white, draw=black] (5.9,9) rectangle (6.1,10.5); 
				\draw[dashed] (4.25,9) -- (7.75,9) node[right] {$z_1$};
				\draw[dashed] (4.25,10.5) -- (7.75,10.5) node[right] {$z_2$};
				\node[above] at (10.5,11) {\textbf{Case \rom{3}}};
				\fill[fill=blue!60!white, draw=black] (9,8) rectangle (10.5,11); 
				\fill[fill=blue!40!white, draw=black] (9,9) rectangle (10.5,10.5); 
				\fill[fill=blue!20!white, draw=black] (9,10.5) rectangle (10.5,11); 
				\fill[fill=blue!60!white, draw=black] (10.5,8) rectangle (12,9); 
				\fill[fill=blue!40!white, draw=black] (10.5,9) rectangle (12,10); 
				\draw[thick, ->] (10.5,8) -- (12.25,8) node[right] {$x$};
				\draw[thick] (10.5,8) -- (10.5,10.5);
				\fill[fill=black!25!white, draw=black] (10.4,9) rectangle (10.6,10.5); 
				\draw[dashed] (8.75,9) -- (12.25,9) node[right] {$z_1$};
				\draw[dashed] (8.75,10.5) -- (12.25,10.5) node[right] {$z_2$};
			\end{tikzpicture}
			\caption{Definition of layers for cells adjacent to a structure interface. $z_1$ and $z_2$ represent the elevation of the base and cover of the structure.}
			\label{fig:Gate Cases}
		\end{figure}
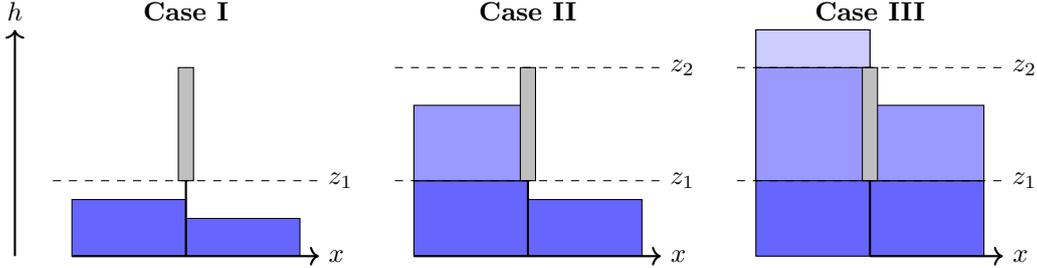
	
		The next step is to assign velocity to the layers. This is achieved by using the following assumptions:
		\begin{itemize}
			\item $hu = \sum_{1}^{n}h_nu_n$ where $h$ is the cell average depth, $u$ is the cell average velocity and $h_n$ and $u_n$ are the depth and velocity of the $n$th layer.
			\item The fluid velocity is exactly equal to zero at the interface/structure for a closed layer.
			\item The fluid velocity in closed layers is much smaller than the fluid velocity in open layers, provided the flow isn't stationary.
		\end{itemize}
		The first assumption equates to the conservation of momentum for the cell following layer velocity assignment, which is a necessary requirement. The second and third assumptions are used to justify assignment of zero velocity to the closed layers. In reality, this results in an underestimation of the fluid velocity for closed layers, which is expected to be small but non-zero, and a subsequent overestimation of the fluid velocity for the open layers. However, this only impacts the computation of the numerical fluxes for the open layers since solution of the Riemann problem for closed layers is dependent on the assumed zero velocity at the interface, which is valid provided the structure is motionless. Furthermore, a more sophisticated discretisation of the layer velocities would likely improve the method, however, this simplified approach has been found to be satisfactory in capturing the approximate vertical velocity profile. Using these assumptions, the velocity for each layer can be determined as shown in Figure \ref{fig:Layers}. Another simplistic velocity assignment approach that can be considered is to assume that the velocity in each of the layers is equal to the cell average velocity. However, this approach was found to produce a significantly underestimated momentum flux as a result of underestimating the flow velocity, especially for the base layer. 
		\begin{figure}[hbt!]
			\centering
			\begin{tikzpicture}
				\node[above] at (3.5,11) {\textbf{Case \rom{2}}};
				\fill[fill=blue!60!white, draw=black] (2,8) rectangle (3.5,9); 
				\draw[<->] (1.5,8) -- (1.5,9);
				\node[left] at (1.5,8.5) {$h_{1,L}$};
				\node[] at (2.75,8.5) {$u_{1,L}$};
				\fill[fill=blue!40!white, draw=black] (2,9) rectangle (3.5,10); 
				\draw[dashed] (1.5,10) -- (3.5,10);
				\draw[<->] (1.5,9) -- (1.5,10);
				\node[left] at (1.5,9.5) {$h_{2,L}$};
				\node[] at (2.75,9.5) {$u_{2,L}$};
				\fill[fill=blue!60!white, draw=black] (3.5,8) rectangle (5,8.75); 
				\draw[dashed] (3.5,8.75) -- (5.5,8.75);
				\draw[<->] (5.5,8) -- (5.5,8.75);
				\node[right] at (5.5,8.375) {$h_{1,R}$};
				\node[] at (4.25,8.375) {$u_{1,R}$};
				\draw[thick, ->] (2,8) -- (5.25,8);
				\draw[thick] (3.5,8) -- (3.5,10.5);
				\fill[fill=black!25!white, draw=black] (3.4,9) rectangle (3.6,10.5); 
				\draw[dashed] (1.75,9) -- (5.25,9);
				\draw[dashed] (1.75,10.5) -- (5.25,10.5);
				\node[above] at (9.5,11) {\textbf{Case \rom{3}}};
				\fill[fill=blue!60!white, draw=black] (8,8) rectangle (9.5,9); 
				\draw[<->] (7.5,8) -- (7.5,9);
				\node[left] at (7.5,8.5) {$h_{1,L}$};
				\node[] at (8.75,8.5) {$u_{1,L}$};
				\fill[fill=blue!40!white, draw=black] (8,9) rectangle (9.5,10.5); 
				\draw[<->] (7.5,9) -- (7.5,10.5);
				\node[left] at (7.5,9.75) {$h_{2,L}$};
				\node[] at (8.75,9.75) {$u_{2,L}$};
				\fill[fill=blue!20!white, draw=black] (8,10.5) rectangle (9.5,11); 
				\draw[<->] (7.5,10.5) -- (7.5,11);
				\node[left] at (7.5,10.75) {$h_{3,L}$};
				\node[] at (8.75,10.7) {$u_{3,L}$};
				\fill[fill=blue!60!white, draw=black] (9.5,8) rectangle (11,9); 
				\draw[<->] (11.5,8) -- (11.5,9);
				\node[right] at (11.5,8.5) {$h_{1,R}$};
				\node[] at (10.25,8.5) {$u_{1,R}$};
				\fill[fill=blue!40!white, draw=black] (9.5,9) rectangle (11,10); 
				\draw[dashed] (9.5,10) -- (11.25,10);
				\draw[<->] (11.5,9) -- (11.5,10);
				\node[right] at (11.5,9.5) {$h_{2,R}$};
				\node[] at (10.25,9.5) {$u_{2,R}$};
				\draw[thick, ->] (8,8) -- (11.25,8);
				\draw[thick] (9.5,8) -- (9.5,10.5);
				\fill[fill=black!25!white, draw=black] (9.4,9) rectangle (9.6,10.5); 
				\draw[dashed] (7.75,9) -- (11.25,9);
				\draw[dashed] (7.75,10.5) -- (11.25,10.5);
				\node[above] at (3.5,7.15) {$h_Lu_L = h_{1,L}u_{1,L} + h_{2,L}u_{2,L}$};
				\node[above] at (3.5,6.5) {$u_{2,L} = 0$};
				\node[above] at (3.5,5.85) {$u_{1,L} = \frac{h_Lu_L}{h_{1,L}}$};
				\node[above] at (9.5,7.15) {$h_Lu_L = h_{1,L}u_{1,L} + h_{2,L}u_{2,L} + h_{3,L}u_{3,L}$};
				\node[above] at (9.5,6.5) {$u_{2,L} = 0 \textrm{ and } u_{open}^L = u_{1,L} = u_{3,L}$};
				\node[above] at (9.5,5.85) {$u_{open}^L = \frac{h_Lu_L}{h_{1,L}+h_{3,L}}$};
			\end{tikzpicture}
			\caption{Velocity assignment with example calculations shown below. $h_L = $ the depth of the non-split left cell, $u_L = $ the cell average velocity, $h_{n,L} = $ the depth of the $n$th split layer, $u_{n,L} = $ the velocity of the $n$th split layer.}
			\label{fig:Layers} 
		\end{figure}
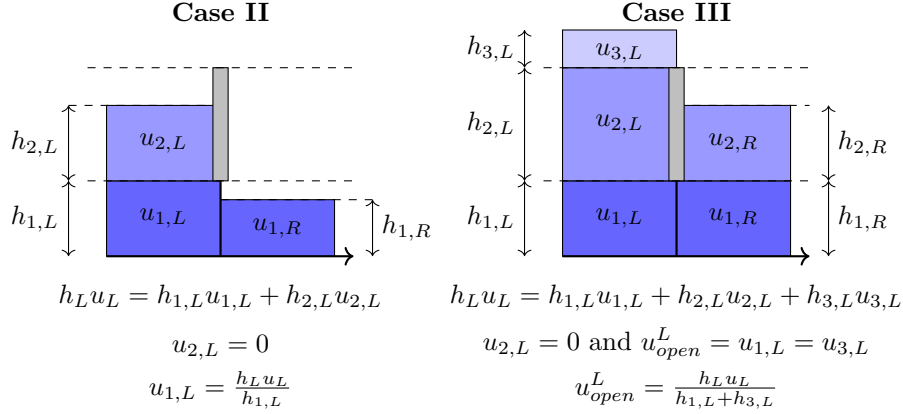
	
		Once the layers and their respective properties are defined the next step is to calculate fluxes layer by layer. As shown by Spinewine et al. \cite{RN39} , due to the lack of coupling between the layers in \eqref{eq: 2-layer SWE 2}, the flow variables in the upper layer are constant across the waves of the bottom layer for the solution of the homogeneous conservation law. This principle can be used to apply the two-layer equations to an $n$-layer system by taking the lower layer to be the layer in question and the upper layer to be the combined layers of flow above. 
		\vspace{5mm} \\ 
		For the closed layers, since the left and right states are separated by a reflective boundary, two fluxes, as opposed to a single flux for the open layers, are to be calculated corresponding to the solution of two Riemann problems involving the left/right layer state and a reflective boundary condition. Once the component fluxes have been calculated for each layer a flux for the left side of the interface can be calculated by summing the left side fluxes for the closed layers with the fluxes for the open layers and a flux for the right side of the interface can be calculated by summing the right side fluxes for the closed layers with the fluxes for the open layers. Updating of the conserved variables is conducted on a cell average basis using these fluxes, meaning that redefinition of the layers and redistribution of the layer properties must be conducted at each timestep based on the updated cell average properties. 
		\begin{algorithm}\label{Alg: 1}
			\caption{Structure cell flux computation algorithm. Subscript $k$ refers to the $k$th layer under consideration. Subscript $u$ refers to the depth of water above the $k$th layer.}
			\LinesNumbered
			\For{each open layer}{
				calculate wavespeeds\;
				\uIf{wet cells}{
					\begin{equation}S_L = u_{open,L} - a_{k,L}q_{k,L} \quad \textrm{ , } \quad S_R = u_{open,R} + a_{k,R}q_{k,R} \tag{1} \end{equation}
					$$a_{k} = \sqrt{(h_{k}+h_{u})g} $$
					$$q_k = \begin{cases} 
						\sqrt{\frac{1}{2}\left[\frac{(h_*+h_{k})h_*}{h_{k}^2} \right]} \textrm{ if } h_* > h_{k} \\[6pt]
						1 \textrm{ if } h_* \leq h_{k}
					\end{cases} $$
					$$h_* = \frac{1}{2}(h_{k,L}+h_{k,R})+\frac{1}{4}\frac{(u_{k,R}-u_{k,L})(h_{k,L}+h_{k,R})}{(a_{k,L}+a_{k,R})} $$
				}
				\uElseIf{left dry cell}{ 
					\begin{equation} 
						S_L =  u_{open,R} - 2a_{k,R} \quad \textrm{ , } \quad S_R =  u_{open,R} + a_{k,R} \tag{2} 
					\end{equation}
				}		
				\uElse{right dry cell; 
					\begin{equation} 
						S_L =  u_{open,L} - a_{k,L} \quad \textrm{ , } \quad S_R = u_{open,L} + 2a_{k,L} \tag{3} 
					\end{equation}
				}
				calculate layer flux\; 
				\begin{equation} \textbf{F}_{k} = \begin{bmatrix}
						h_{k}u_{k} \\
						\frac{(h_{k}u_{k})^2}{h_{k}} + \frac{1}{2}gh_{k}^2 + \chi gh_{u}h_{k} \tag{4} \\ 
				\end{bmatrix}  \end{equation}
				\begin{equation} \textbf{F}_{layer} = \begin{cases}
						\textbf{F}_{k,L} \textrm{ if } S_L > 0 \\[6pt]
						\textbf{F}^{hll} = \frac{S_R\textbf{F}_{k,L}-S_L\textbf{F}_{k,R}+S_RS_L(\textbf{U}_R-\textbf{U}_L)}{S_R-S_L} \textrm{ if } S_L \leq 0 \leq S_R \\[6pt] \tag{5}
						\textbf{F}_{k,R} \textrm{ if } S_R < 0
				\end{cases}  \end{equation}
			}
			\For{each closed layer}{ 
				\For{the left side}{ 
					introduce fictitious ghost cell for right state (equal depth and equal and opposite velocity)\; 
					calculate wavespeeds using $(1)$\; 
					calculate layer flux using $(4) \& (5)$\; 
				}
				\For{the right side}{ 
					introduce fictitious ghost cell for left state (equal depth and equal and opposite velocity)\; 
					calculate wavespeeds using $(1)$\; 
					calculate layer flux using $(4) \& (5)$\; 
				}
			}
		\end{algorithm}
		\vspace{5mm} \\
		The outlined procedure, which is summarised in the pseudocode contained within Algorithm 1, can be demonstrated by considering the scenarios presented in Cases \rom{1}-\rom{3}. Case \rom{1} is a trivial case, requiring no special treatment in comparison to the non-structure cells since there is only one open layer containing depth of flow for both left and right states. Case \rom{2} is more complex, requiring the resolution of two fluxes via the solution of two Riemann problems: a flux for the base layer and a flux for the left side of the structure. As illustrated in Figure \ref{fig:Riemann Problem 1}, the solutions to the Riemann problems constructed at the left and right side of the interface for a closed layer always produce a depth flux of zero and a momentum flux equal to the pressure force exerted by the layer of water on the wall (assuming a hydrostatic pressure distribution), which is equal to $1/2gh^2_{2,L}$ for Case \rom{2}). 
		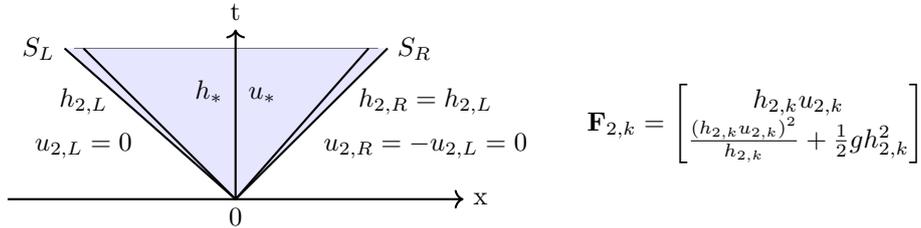
\begin{figure}[hbt!]
			\centering
			\begin{tikzpicture}
				\node [] (A) at ( 3, 0) {};
				\node [] (B) at (0.75, 2) {};
				\node [] (C) at (5, 2) {};
				\draw (A) -- (B) -- (C) -- (A);
				\begin{scope} 
					\fill [blue!10!white] (A.center) -- (B.center) -- (C.center) -- cycle;
				\end{scope}
				\draw[thick,->] (0,0) -- (6,0) node[right] {x}; 
				\draw[thick,->] (3,0) -- (3,2.25) node[above] {t};
				\node[below] at (3,0) {0};
				\draw[thick] (3,0) -- (1,2);
				\draw[thick] (3,0) -- (0.75,2) node[left] {$S_L$};
				\draw[thick] (3,0) -- (5,2) node[right] {$S_R$};
				\draw[thick] (3,0) -- (4.75,2);
				\node[above] at (1,1) {$h_{2,L}$};
				\node[below] at (1,1) {$u_{2,L}=0$};
				\node[above] at (5.5,1) {$h_{2,R}=h_{2,L}$};
				\node[below] at (5.5,1) {$u_{2,R}=-u_{2,L}=0$};
				\node[above] at (2.65,1.15) {$h_*$};
				\node[above] at (3.35,1.15) {$u_*$};
				\node[right] at (7.5,1) {$\textbf{F}_{2,k} = \begin{bmatrix}
						h_{2,k}u_{2,k} \\
						\frac{(h_{2,k}u_{2,k})^2}{h_{2,k}} + \frac{1}{2}gh_{2,k}^2 \\
					\end{bmatrix}$};
			\end{tikzpicture}
			\caption{Structure of the general solution of the Riemann problem for the closed layer in the left cell for Case \rom{2}.}
			\label{fig:Riemann Problem 1}
		\end{figure}
	
		Case \rom{3} introduces further complexity, requiring the resolution of four fluxes via the solution of four Riemann problems: a flux for the base layer, a flux for the left side of the structure interface, a flux for the right side of the structure interface and a flux for the uppermost layer. The Riemann problem constructed for the uppermost layer includes a dry right bed as there is no depth of flow in this layer for the right state. As a consequence, the wave pattern and corresponding wavespeeds displayed in Figure \ref{fig:Wet-Dry Bed Wavepatterns} are utilised. An alternative approach is to model these types of layers as free outfalls, imposing a critical depth condition in the left or right state which contains no depth of flow.
		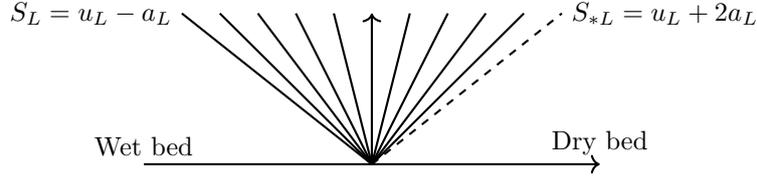
\begin{figure}[hbt!]
			\centering
			\begin{tikzpicture}
				\draw[thick,->] (6,8) -- (6,10);
				\draw[thick] (6,8) -- (3.5,10) node[left] {$S_{L}=u_L-a_L$};
				\draw[thick] (6,8) -- (4,10);
				\draw[thick] (6,8) -- (4.5,10);
				\draw[thick] (6,8) -- (5,10);
				\draw[thick] (6,8) -- (5.5,10);
				\draw[thick] (6,8) -- (6.5,10);
				\draw[thick] (6,8) -- (7,10);
				\draw[thick] (6,8) -- (7.5,10);
				\draw[thick] (6,8) -- (8,10);
				\draw[thick, dashed] (6,8) -- (8.5,10) node[right] {$S_{*L}=u_L+2a_L$};
				\node[above] at (3,8) {Wet bed};
				\node[above] at (9,8) {Dry bed};
				\draw[thick,->] (3,8) -- (9,8);
			\end{tikzpicture}
			\caption{Wave pattern for the one-dimensional case with a right dry bed. $a_k=\sqrt{gh_k}$.}
			\label{fig:Wet-Dry Bed Wavepatterns}
		\end{figure}
		Figure \ref{fig:Structure Cell Updating} illustrates the process used to update the conserved variables for the structure cells following summation of the component fluxes. $\textbf{F}^{(-)}$, the flux for the left side of the interface, is determined by summing the component fluxes for the open layers with the component fluxes for the closed layers on the left side of the interface. Likewise, $\textbf{F}^{(+)}$, the flux for the right side of the interface, is determined by summing the component fluxes for the open layers with the component fluxes for the closed layers on the right side of the interface. In the case that $\textbf{F}^{(-)} \neq \textbf{F}^{(+)}$, which occurs when the left and right states are not in equilibrium, there is a loss of momentum from the shallow water system at each timestep equal to $\Delta t/\Delta x(\textbf{F}^{(+)} - \textbf{F}^{(-)})$, which is equal to the resultant hydrostatic pressure force exerted on the structure multiplied by the ratio of the timestep to the cell size. The resultant hydrostatic pressure force is valuable for applications concerned with determining the structural failure of hydraulic structures due to fluid-structure interactions.
		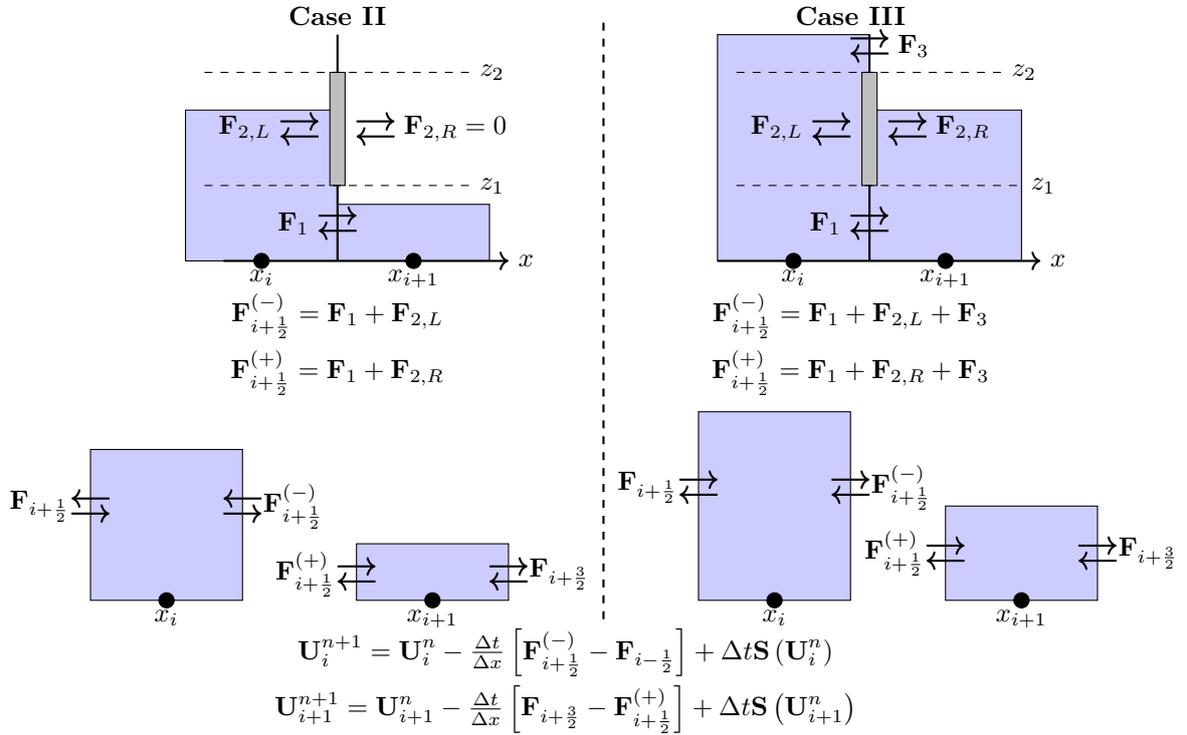
\begin{figure}[hbt!]
			\centering
			\begin{tikzpicture}
				\node[above] at (3,3) {\textbf{Case \rom{2}}};
				\fill[fill=blue!20!white, draw=black] (1,0) rectangle (3,2); 
				\fill[fill=black, draw=black] (2,0) circle (0.1cm); 
				\node[below] at (2,0) {$x_{i}$};
				\fill[fill=blue!20!white, draw=black] (3,0) rectangle (5,0.75); 
				\fill[fill=black, draw=black] (4,0) circle (0.1cm); 
				\node[below] at (4,0) {$x_{i+1}$};
				\draw[thick, ->] (1.5,0) -- (5.25,0) node[right] {$x$};
				\draw[thick] (3,0) -- (3,3);
				\fill[fill=black!25!white, draw=black] (2.9,1) rectangle (3.1,2.5); 
				\draw[dashed] (1.25,1) -- (4.75,1) node[right] {$z_1$};
				\draw[dashed] (1.25,2.5) -- (4.75,2.5) node[right] {$z_2$};
				\draw[thick, ->] (2.75,0.6) -- (3.25,0.6); 
				\draw[thick, ->] (3.25,0.4) -- (2.75,0.4); 
				\node[left] at (2.75,0.5) {$\textbf{F}_{1}$};
				\draw[thick, ->] (2.25,1.85) -- (2.75,1.85); 
				\draw[thick, ->] (2.75,1.65) -- (2.25,1.65); 
				\node[left] at (2.25,1.75) {$\textbf{F}_{2,L}$};
				\draw[thick, ->] (3.25,1.85) -- (3.75,1.85); 
				\draw[thick, ->] (3.75,1.65) -- (3.25,1.65); 
				\node[right] at (3.75,1.75) {$\textbf{F}_{2,R}=0$};
				
				\draw[thick, dashed] (6.5,-4.75) -- (6.5,3.25); 
				
				\fill[fill=blue!20!white, draw=black] (-0.25,-4.5) rectangle (1.75,-2.5); 
				\fill[fill=black, draw=black] (0.75,-4.5) circle (0.1cm); 
				\node[below] at (0.75,-4.5) {$x_{i}$};
				\node[right] at (1.9,-3.25) {$\textbf{F}_{i+\frac{1}{2}}^{(-)}$};
				\draw[thick, ->] (1.5,-3.35) -- (2,-3.35); 
				\draw[thick, ->] (2,-3.15) -- (1.5,-3.15); 
				\node[left] at (-0.4,-3.25) {$\textbf{F}_{i+\frac{1}{2}}$};
				\draw[thick, ->] (-0.5,-3.35) -- (0,-3.35); 
				\draw[thick, ->] (0,-3.15) -- (-0.5,-3.15); 
				\fill[fill=blue!20!white, draw=black] (3.25,-4.5) rectangle (5.25,-3.75); 
				\node[left] at (3.1,-4.15) {$\textbf{F}_{i+\frac{1}{2}}^{(+)}$};
				\draw[thick, ->] (3,-4.05) -- (3.5,-4.05); 
				\draw[thick, ->] (3.5,-4.25) -- (3,-4.25); 
				\node[right] at (5.4,-4.15) {$\textbf{F}_{i+\frac{3}{2}}$};
				\draw[thick, ->] (5,-4.05) -- (5.5,-4.05); 
				\draw[thick, ->] (5.5,-4.25) -- (5,-4.25); 
				\fill[fill=black, draw=black] (4.25,-4.5) circle (0.1cm); 
				\node[below] at (4.25,-4.5) {$x_{i+1}$};
				
				\fill[fill=blue!20!white, draw=black] (7.75,-4.5) rectangle (9.75,-2); 
				\fill[fill=black, draw=black] (8.75,-4.5) circle (0.1cm); 
				\node[below] at (8.75,-4.5) {$x_{i}$};
				\node[right] at (9.9,-3) {$\textbf{F}_{i+\frac{1}{2}}^{(-)}$};
				\draw[thick, ->] (9.5,-2.9) -- (10,-2.9); 
				\draw[thick, ->] (10,-3.1) -- (9.5,-3.1); 
				\node[left] at (7.6,-3) {$\textbf{F}_{i+\frac{1}{2}}$};
				\draw[thick, ->] (7.5,-2.9) -- (8,-2.9); 
				\draw[thick, ->] (8,-3.1) -- (7.5,-3.1); 
				\fill[fill=blue!20!white, draw=black] (11,-4.5) rectangle (13,-3.25); 
				\node[left] at (10.85,-3.875) {$\textbf{F}_{i+\frac{1}{2}}^{(+)}$};
				\draw[thick, ->] (10.75,-3.775) -- (11.25,-3.775); 
				\draw[thick, ->] (11.25,-3.975) -- (10.75,-3.975); 
				\node[right] at (13.15,-3.875) {$\textbf{F}_{i+\frac{3}{2}}$};
				\draw[thick, ->] (12.75,-3.775) -- (13.25,-3.775); 
				\draw[thick, ->] (13.25,-3.975) -- (12.75,-3.975); 
				\fill[fill=black, draw=black] (12,-4.5) circle (0.1cm); 
				\node[below] at (12,-4.5) {$x_{i+1}$};
				
				\node[above] at (3,-1.15) {$\textbf{F}_{i+\frac{1}{2}}^{(-)} = \textbf{F}_1 + \textbf{F}_{2,L}$};
				\node[above] at (3,-1.9) {$\textbf{F}_{i+\frac{1}{2}}^{(+)} = \textbf{F}_1 + \textbf{F}_{2,R}$};
				
				\node[above] at (9.75,-1.15) {$\textbf{F}_{i+\frac{1}{2}}^{(-)} = \textbf{F}_1 + \textbf{F}_{2,L} + \textbf{F}_3$};
				\node[above] at (9.75,-1.9) {$\textbf{F}_{i+\frac{1}{2}}^{(+)} = \textbf{F}_1 + \textbf{F}_{2,R} + \textbf{F}_3$};
				
				\node[above] at (9.75,3) {\textbf{Case \rom{3}}};
				\fill[fill=blue!20!white, draw=black] (8,0) rectangle (10,3); 
				\fill[fill=black, draw=black] (9,0) circle (0.1cm); 
				\node[below] at (9,0) {$x_{i}$};
				\fill[fill=blue!20!white, draw=black] (10,0) rectangle (12,2); 
				\fill[fill=black, draw=black] (11,0) circle (0.1cm); 
				\node[below] at (11,0) {$x_{i+1}$};
				\draw[thick, ->] (8,0) -- (12.25,0) node[right] {$x$};
				\draw[thick] (10,0) -- (10,3);
				\fill[fill=black!25!white, draw=black] (9.9,1) rectangle (10.1,2.5); 
				\draw[dashed] (8.25,1) -- (12,1) node[right] {$z_1$};
				\draw[dashed] (8.25,2.5) -- (11.75,2.5) node[right] {$z_2$};
				\draw[thick, ->] (9.75,0.6) -- (10.25,0.6); 
				\draw[thick, ->] (10.25,0.4) -- (9.75,0.4); 
				\node[left] at (9.75,0.5) {$\textbf{F}_{1}$};
				\draw[thick, ->] (9.25,1.85) -- (9.75,1.85); 
				\draw[thick, ->] (9.75,1.65) -- (9.25,1.65); 
				\node[left] at (9.25,1.75) {$\textbf{F}_{2,L}$};
				\draw[thick, ->] (10.25,1.85) -- (10.75,1.85); 
				\draw[thick, ->] (10.75,1.65) -- (10.25,1.65); 
				\node[right] at (10.75,1.75) {$\textbf{F}_{2,R}$};
				\draw[thick, ->] (9.75,2.95) -- (10.25,2.95); 
				\draw[thick, ->] (10.25,2.75) -- (9.75,2.75); 
				\node[right] at (10.25,2.85) {$\textbf{F}_{3}$};
				
				\node[below] at (6,-4.75) {$\textbf{U}^{n+1}_i = \textbf{U}^{n}_i - \frac{\Delta t}{\Delta x}\left[\textbf{F}_{i+\frac{1}{2}}^{(-)} -  \textbf{F}_{i-\frac{1}{2}}\right] + \Delta t \textbf{S}\left(\textbf{U}_i^n \right)$};
				\node[below] at (6,-5.5) {$\textbf{U}^{n+1}_{i+1} = \textbf{U}^{n}_{i+1} - \frac{\Delta t}{\Delta x}\left[\textbf{F}_{i+\frac{3}{2}} -  \textbf{F}_{i+\frac{1}{2}}^{(+)}\right] + \Delta t \textbf{S}\left(\textbf{U}_{i+1}^n \right)$};
			\end{tikzpicture}
			\caption{Updating of the left and right structure cells following summation of the component fluxes at the structure interface.}
			\label{fig:Structure Cell Updating}
		\end{figure}
		\section{Model Validation}
		\begin{figure}[hbt!]
			\begin{center}
					\includegraphics[width=.6\linewidth]{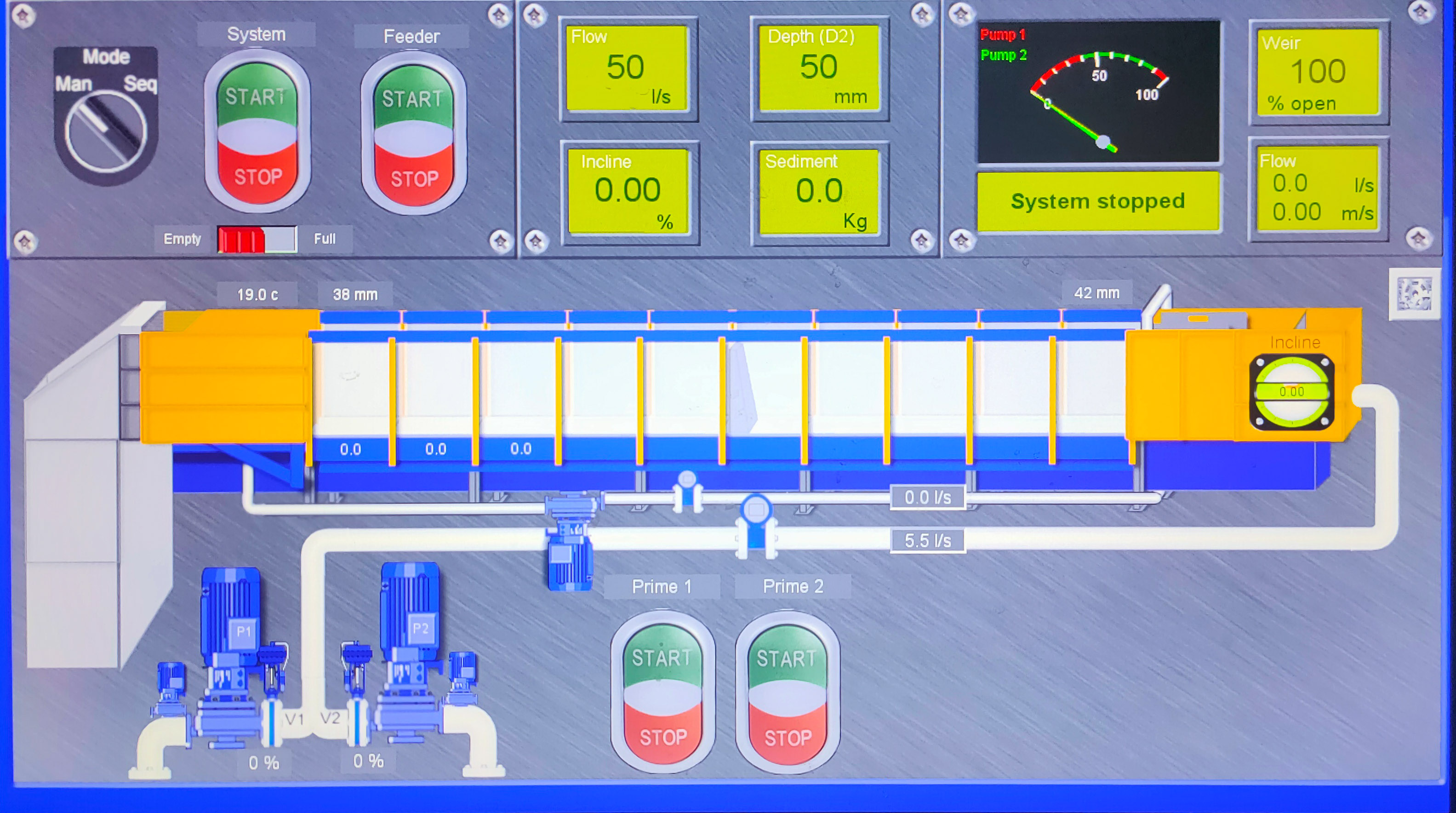}
			\end{center}
			\caption{Control panel for the S100 Research Flume, including a schematic of the flume. Two pumps are used to draw water from the sump, supplying a constant flow rate to the upstream (right) end of the flume. The flow rate is measured using an electromagnetic flow meter.}
			\label{Fig: Flume Control Panel}
		\end{figure}
		In order to validate the method outlined in the previous section, numerical results are compared to steady state measurements taken from experiments conducted in Newcastle University's Armfield S100 Research Flume. As shown in Figure \ref{Fig: Flume Control Panel}, the S100 Research Flume is a $12.5m$ long, $1m$ wide, $0.8m$ deep flume capable of producing flow rates up to $400ls^{-1}$. At $5m$ downstream, grooves in the walls of the flume enable barriers to be slotted into the cross-section enabling their effect to be studied. A series of tests were conducted in the flume using a range of flows and barrier geometries. In order to simplify the numerical scheme the flume was set to zero tilt for all tests. The basic test conditions are outlined in Table \ref{Table: Test Conditions}. The full validation dataset is available as supplementary material for potential future use by other researchers. 
		\begin{table}[ht]
			\begin{tabular}{ |c|c|c|c|c| }
				\hline
				\multicolumn{5}{|c|}{Validation Test Cases} \\
				\hline
				Test Case & $q \textrm{ } (ls^{-1})$ & $z_1 \textrm{ } (mm)$ & $z_2  \textrm{ } (mm)$ & Description \\
				\hline
				Test $1$ & $130$ & $116$ & $316$ & Flow under barrier. \\
				Test $2$ & $130$ & $122$ & $322$ & Flow under barrier. \\
				Test $3$ & $35$ & $32$ & $232$ & Flow under barrier. \\
				Test $4$ & $130$ & $32$ & $232$ & Overtopped barrier. \\
				Test $5$ & $24$ & $32$ & $232$ & Flow under barrier. \\
				Test $6$ & $20$ & $32$ & $232$ & Flow under barrier. \\
				Test $7$ & $275$ & $116$ & $316$ &Overtopped barrier. \\
				Test $8$ & $175$ & $116$ & $316$ & Overtopped barrier. \\
				Test $9$ & $177$ & $105$ & $405$ & Overtopped barrier. \\
				Test $10$ & $150$ & $105$ & $405$ & Flow under barrier. \\
				Test $11$ & $225$ & $116$ & $316$ & Overtopped barrier. \\
				Test $12$ & $19$ & $24$ & $324$ & Flow under barrier. \\
				\hline
			\end{tabular}
			\centering
			\caption{Validation test conditions. $q$ is the flow rate, $z_1$ is the elevation of the base of the barrier above the flume bed and $z_2$ is the elevation of the cover of the barrier.}
			\label{Table: Test Conditions}
		\end{table}
		\subsection{Numerical Setup}
		All numerical tests were conducted on a $12.5m$ 1D spatial domain, discretised into a structured grid comprised of $0.01m$ cells. In order to ensure satisfaction of the Courant-Friedrichs-Lewy condition a Courant number of $C = (0.95\Delta x)/(S^n_{max})$ was used to determine a stable timestep, where $S_{max}^n$ is the maximum absolute wave speed at time level $n$. Since the bed slope is set to $0\%$ this has the intended effect of simplifying the source terms, only requiring the friction source term to be resolved, facilitating clearer analysis of the novel solution procedure. The friction source term is resolved using a splitting method presented by Liang and Marche \cite{RN27}:
		\begin{equation}\label{eq: Friction Source Term} \notag
			q^{n+1} = q^{n} - \Delta t S^n_c = q^{n} - \Delta t \left(\frac{\tau_f}{1 + \Delta t\frac{\partial \tau_f}{\partial q}} \right)^n = q^{n} - \Delta t \left(\frac{Cu|u|}{1 + \frac{2\Delta tC_f|q|}{h^2}} \right)^n 
		\end{equation}
		The following simple limiter is also implemented to ensure stability in regions where the water depth approaches zero:
		\begin{equation}
			S_c^n = \frac{q^n}{\Delta t} \textrm{ if } |\Delta t S_c^n| > |q^n|
		\end{equation}
		A Manning's n of $0.012$ is assumed for all test cases. 
		\subsubsection{Upstream Boundary Condition}
		\begin{figure}[hbt!]
			\centering
			\begin{tikzpicture}
				\draw[thick] (0,0) -- (12,0); 
				\draw[dashed] (0,0) -- (0,2);
				\fill[fill=black!25!white, draw=black] (0,0) rectangle (2,1.5);
				\draw[thick] (2,0) -- (2,2.25) node[above] {$a$};
				\draw[thick] (2,0) -- (2,-0.5) node[below] {$x_{\frac{1}{2}}$};
				\node[] at (1,0.75) {\LARGE$\textbf{U}_{-1}$};
				\fill[fill=blue!25!white, draw=black] (2,0) rectangle (4,1.5);
				\draw[dashed] (4,0) -- (4,2);
				\node[] at (3,0.75) {\LARGE$\textbf{U}_{0}$};
				\draw[dashed] (6,0) -- (6,2);
				\node[] at (5,0.75) {\LARGE...};
				\draw[dashed] (8,0) -- (8,2);
				\node[] at (7,0.75) {\LARGE...};
				\fill[fill=blue!25!white, draw=black] (8,0) rectangle (10,1.5);
				\draw[thick] (10,0) -- (10,2.25) node[above] {$b$};
				\draw[thick] (10,0) -- (10,-0.5) node[below] {$x_{N+\frac{1}{2}}$};
				\node[] at (9,0.75) {\LARGE$\textbf{U}_{N}$};
				\fill[fill=black!25!white, draw=black] (10,0) rectangle (12,1.5);
				\draw[dashed] (12,0) -- (12,2);
				\node[] at (11,0.75) {\LARGE$\textbf{U}_{N+1}$};
			\end{tikzpicture}
			\caption{Computational domain $[a,b]$, with exterior ghost cells for specifying boundary conditions.}
			\label{fig:Ghost Cells 1}
		\end{figure}
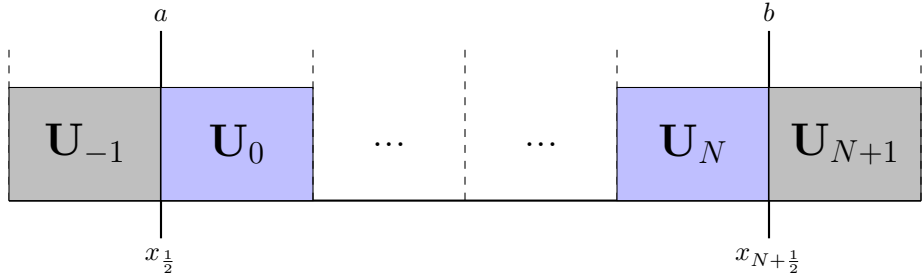
		An upstream inflow boundary condition is implemented to ensure a constant flow rate is maintained through the upstream boundary, replicating the pump system used by the S100 flume.
		This is achieved using exterior ghost cells as shown in Figure \ref{fig:Ghost Cells 1}; consider the left boundary $\textbf{U}_{-1}=\textbf{U}_L$ (ghost cell) and $\textbf{U}_{0}=\textbf{U}_R$. Using the Riemann invariant: $$u_L + 2a_L = u_R + 2a_R$$ and substituting:
		\begin{gather*}
			Q_{in} = h_Lu_L \\
			u_L = \frac{Q_{in}}{h_L} \\
			a_k = \sqrt{gh_k}
		\end{gather*}
		gives:
		\begin{gather*}
			\frac{Q_{in}}{h_L} + 2\sqrt{gh_L} = u_R + 2\sqrt{gh_R}
		\end{gather*}
		Which can be rearranged to give the following function:
		\begin{gather*}
			\frac{Q_{in}}{h_L} + 2\sqrt{gh_L} - u_R - 2\sqrt{gh_R} = 0 \\
			\frac{Q_{in}}{h_L} + 2\sqrt{g}(\sqrt{h_L}-\sqrt{h_R}) - u_R  = 0 
		\end{gather*}
		Let:
		\begin{gather*}
			f(h_L, h_R, u_R) = \frac{Q_{in}}{h_L} + 2\sqrt{g}(\sqrt{h_L}-\sqrt{h_R}) - u_R \\
			\frac{\,d}{\,dh_L}f(h_L, h_R, u_R) = -\frac{Q_{in}}{h_L^2} + \frac{\sqrt{g}}{\sqrt{h_L}}
		\end{gather*}
		A suitable numerical method such as the Newton-Raphson method can then be used to determine $h_L$, using an initial guess of $h_0=h_R$:
		\begin{gather*}
			x_{n+1} = x_n - \frac{f(x_n)}{f'(x_n)}
		\end{gather*}
		Finally $u_L$ may be determined using: $$u_L = \frac{Q_{in}}{h_L}$$ providing the initial conditions for use in the Riemann solver at the upstream boundary. Use of the Riemann invariants, as described above, requires that the wave connecting the states at the boundary is a rarefaction wave. This is due to the fact that Riemann invariants are only perfectly invariant across simple waves (the two connected states lie on the same integral curve). The two states connected by a shock wave do not lie on the same integral curve and instead are connected by a Hugoniot locus. Therefore should a shock wave occur at the boundary, which may physically occur when the velocity in the first cell is opposite in direction to the inflow, the outlined method is no longer valid and other conditions must be imposed. However, for the outlined test cases this does not occur so the method is suitable.
		\subsubsection{Downstream Boundary Condition}
		At the downstream end of the S100 flume is a sloped free-outfall as shown in Figure \ref{fig:Structure Cell Updating}. In order to approximate the behaviour of the flow at this boundary a critical depth boundary condition is imposed. This is achieved by using the following initial conditions at the downstream boundary (Figure \ref{fig:Riemann Problem 2}):
		\begin{gather}
			\textbf{U}_{N}^n = \begin{bmatrix}
				h_N^n \\
				h_N^nu_N^n
			\end{bmatrix}
			\textrm{ , } \textbf{U}_{N+1}^n = \begin{bmatrix}
				h_{N+1} = \left(\frac{\left(h_N^nu_N^n\right)^2}{g}\right)^{\frac{1}{3}}   \\
				h_{N+1}^nu_{N+1}^n = h_N^nu_N^n
			\end{bmatrix}
		\end{gather}
		\begin{figure}[hbt!]
			\centering
			\begin{tikzpicture}
				\node [] (A) at ( 3, 0) {};
				\node [] (B) at (0.75, 1.9) {};
				\node [] (C) at (5, 1.9) {};
				\draw (A) -- (B) -- (C) -- (A);
				\begin{scope} 
					\fill [blue!10!white] (A.center) -- (B.center) -- (C.center) -- cycle;
				\end{scope}
				\draw[thick,->] (0,0) -- (6,0) node[right] {x}; 
				\draw[thick,->] (3,0) -- (3,2.15) node[above] {t};
				\node[below] at (3,0) {0};
				\draw[thick] (3,0) -- (1,1.9);
				\draw[thick] (3,0) -- (0.75,1.9) node[above] {$S_L$};
				\draw[thick] (3,0) -- (5,1.9) node[above] {$S_R$};
				\draw[thick] (3,0) -- (4.75,1.9);
				\node[left] at (-0.75,1.15) {$ $};
				\node[above] at (1,0.85) {$h_{L}$};
				\node[below] at (1,0.85) {$u_{L}$};
				\node[above] at (6.25,1) {$h_{R}=\left(\frac{\left(h_Lu_L\right)^2}{g}\right)^{\frac{1}{3}} $};
				\node[below] at (5.5,1) {$u_{R}=\frac{h_Lu_L}{h_R}$};
				\node[above] at (2.65,1) {$h_*$};
				\node[above] at (3.35,1) {$u_*$};
			\end{tikzpicture}
			\caption{Structure of the general solution of the Riemann problem for the downstream boundary with a critical depth condition.}
			\label{fig:Riemann Problem 2}
		\end{figure}
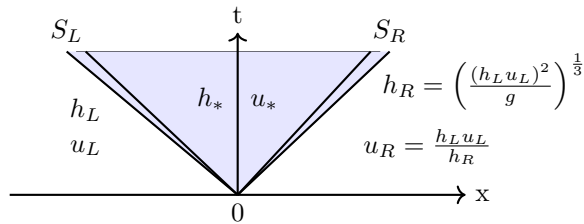
		\subsection{Results}
		The test cases can be categorised into two primary flow conditions: 
		\begin{enumerate}
			\item Flow moving underneath a barrier, which is analogous to flow moving under a gate.
			\item Flow which completely inundates the barrier, analogous to an overtopped bridge structure.
		\end{enumerate}
		Through comparison between the numerical predictions and experimental data for a selection of these test cases, the suitability of the solver for these flow conditions can be determined.
		\subsubsection{Flow Under a Gate}
		\begin{figure}[hbt!]
			\begin{center}
				\includegraphics[width=.95\linewidth]{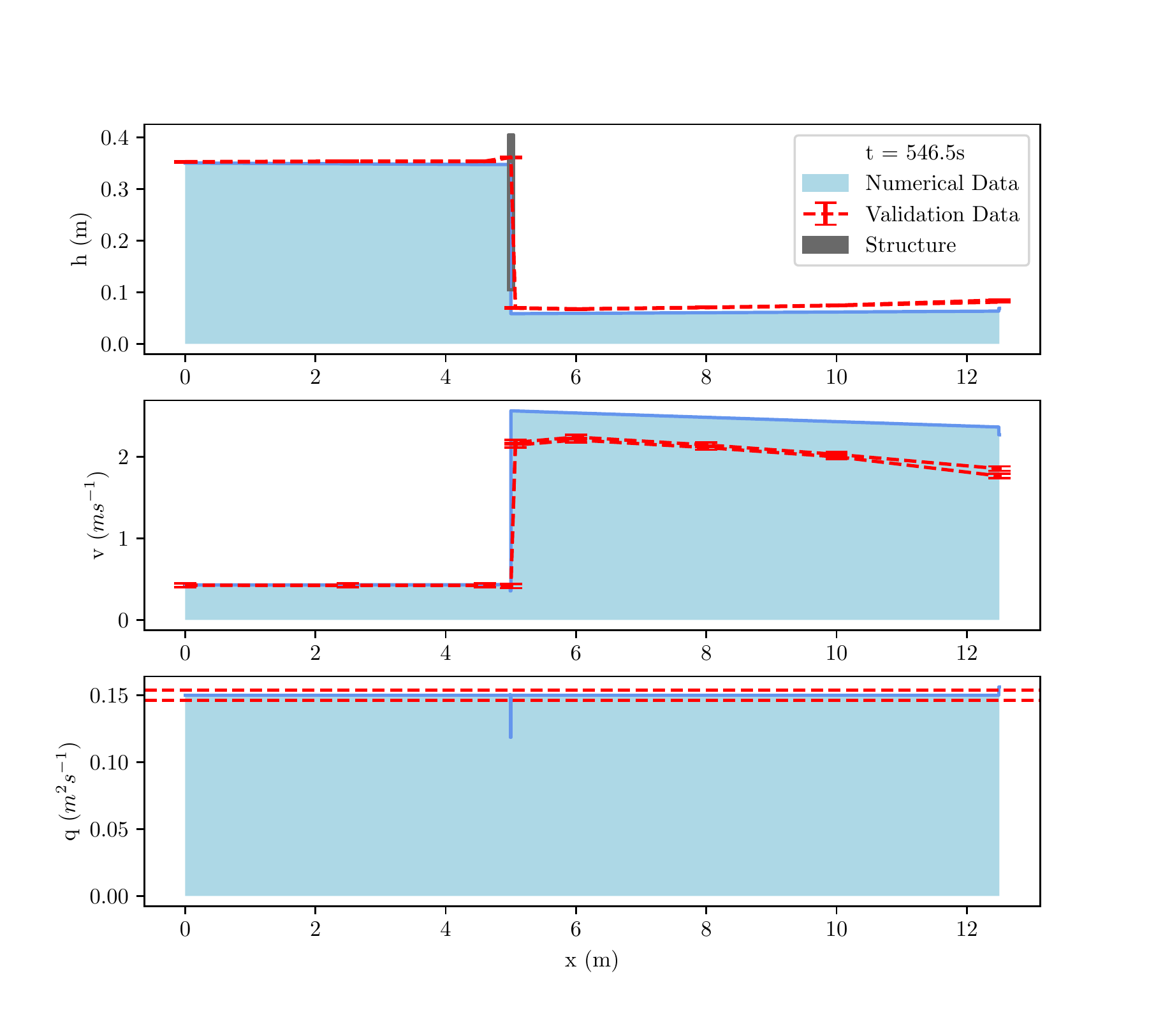}
			\end{center}
			\caption{Comparison between numerical and experimental results for test case $10$.}
			\label{Fig: Test Case Flow Under a Gate 1}
		\end{figure}
		Figures \ref{Fig: Test Case Flow Under a Gate 1} and \ref{Fig: Test Case Flow Under a Gate 2} demonstrate that the solver adequately captures the behaviour induced by flow moving underneath a gate-type structure for a range of flows and barrier geometries. It can be seen that for test case 10, shown in Figure \ref{Fig: Test Case Flow Under a Gate 1}, that the solver contributed to the accurate prediction of flow depths upstream and downstream of the barrier, with a slight overestimation of the velocity and subsequent slight underestimation of the depth downstream of the barrier. For test case 2, shown in Figure \ref{Fig: Test Case Flow Under a Gate 2} there is a larger, albeit still acceptable, underestimation of the depth upstream of the barrier. Downstream of the barrier, the depth prediction is accurate, only minimally deviating from the measured data. For both of the presented test cases, it is likely that the errors are a consequence of a difference in the approximated and real vertical velocity profile at the barrier. It is also worth noting that there is a localised change in $q$ in the immediate cells downstream of the barrier for all test cases. This occurs as a result of the aforementioned $\Delta t/\Delta x(\textbf{F}^{(+)} - \textbf{F}^{(-)})$ term. As noted by Dazzi et al. \cite{RN101}, this is a common feature for schemes attempting to model partial barriers to flow such as Ratia et al. \cite{RN102}, Maranzoni and Mignosa \cite{RN261} and Zhao et al. \cite{RN270}. 
		\begin{figure}[hbt!]
			\begin{center}
				\includegraphics[width=.95\linewidth]{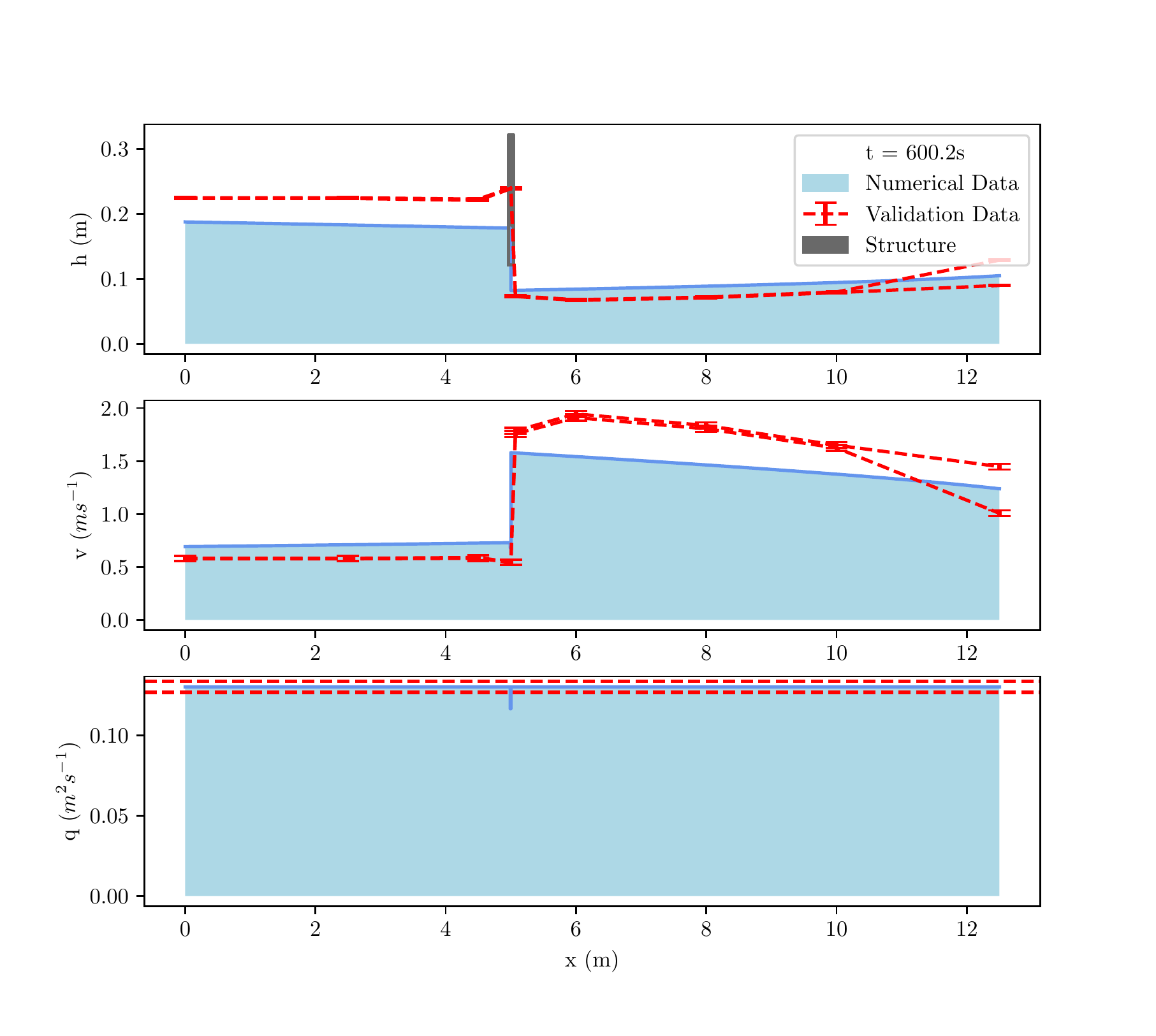}
			\end{center}
			\caption{Comparison between numerical and experimental results for test case $2$.}
			\label{Fig: Test Case Flow Under a Gate 2}
		\end{figure} \clearpage
		
		\subsubsection{Inundated Bridge Structure}
		\begin{figure}[hbt!]
			\begin{center}
				\includegraphics[width=.95\linewidth]{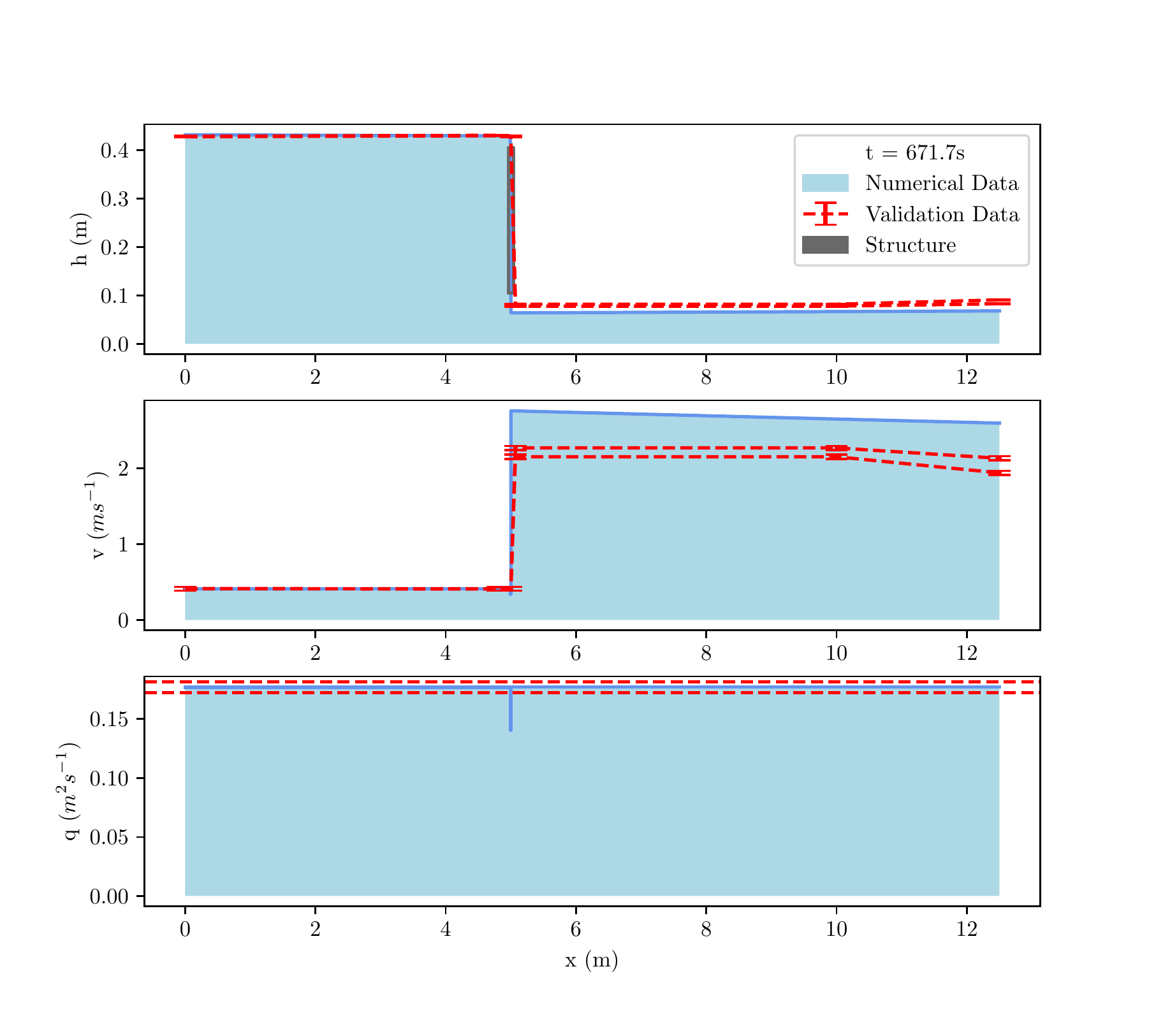}
			\end{center}
			\caption{Comparison between numerical and experimental results for Test case $9$.}
			\label{Fig: Test Case Inundated Gate 1}
		\end{figure}
		Figures \ref{Fig: Test Case Inundated Gate 1} and \ref{Fig: Test Case Inundated Gate 2} demonstrate that the solver is capable of adequately capturing the flow behaviour induced by an overtopped barrier. It can be seen in Figure \ref{Fig: Test Case Inundated Gate 1} that the solver contributed to the accurate prediction of the depth throughout the domain. Similarly, depth predictions for Figure \ref{Fig: Test Case Inundated Gate 2} can be seen to be accurate for a different barrier configuration at the same flow rate. For both of the presented test cases there is a small overestimation of the velocity downstream of the barrier which contributes to an underestimation of the downstream flow depth. As the barrier becomes more significantly overtopped, the predictions downstream of the barrier become less accurate however, the general behaviour is still within an acceptable range in comparison with the experimental data. Once more, this is likely a consequence of the simplistic velocity assignment as well as an increase in vertical motion immediately after the barrier; scenarios where the vertical velocity becomes significant violate the underlying assumptions for the conservation law and the scheme becomes unsuitable. Furthermore, it is important to only consider modelling cases for which the flow behaviour does not substantially violate the underlying assumptions for the conservation law and solution procedure. These primary assumptions include: a hydrostatic pressure distribution, that the pressure is atmospheric at the fluid surface and that the velocity in the vertical direction is negligible.
		\begin{figure}[hbt!]
			\begin{center}
				\includegraphics[width=.95\linewidth]{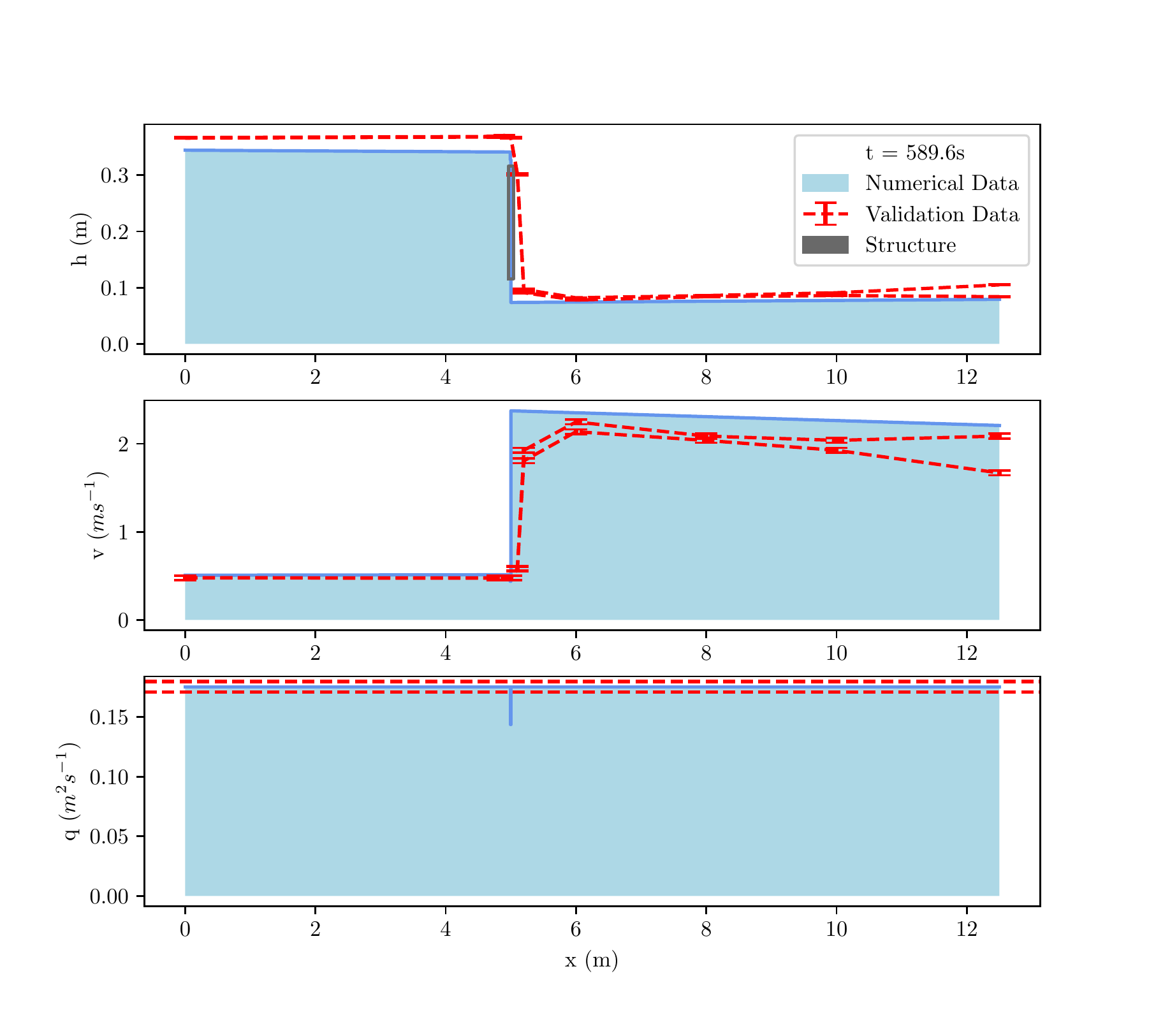}
			\end{center}
			\caption{Comparison between numerical and experimental results for Test case $8$.}
			\label{Fig: Test Case Inundated Gate 2}
		\end{figure} \clearpage
		\section{Conclusion}
		A novel Riemann solver capable of modelling the influence of structures on flood flows within 1D or 2D hydrodynamic flood models has been presented. The validation process demonstrates that the solver is able to adequately capture the flow behaviour for a range of partial barriers to flow at a range of flow rates. Accuracy of the predictions is mostly dependent on how well the velocity assignment process captures the vertical velocity profile at the barrier. As a result, further work to derive a more sophisticated and representative approach to determining the layer velocities for structure cells is proposed. However, in its current state the proposed solver still presents a unique physically based and robust approach to modelling hydraulic structures using a conservative form of the conservation laws. Furthermore, since the solution procedure only applies locally in the region surrounding an interface at which a structure is being modelled, implementation is feasible for both established and developing hydrodynamic models using finite volume schemes to solve the shallow water equations. The biggest barrier to implementation is the availability and resolution of the required data for structures and suitable meshing algorithms. The latter of which is also to be the subject of future work.
			
		Addition of the capability to model a variety of in-channel hydraulic structures such as gates and bridges has the potential to significantly improve the accuracy and flexibility of inundation predictions and therefore contemporary hydrodynamic modelling practice. Further potential applications include but are not limited to:
		\begin{itemize}
			\item Infrastructure resilience modelling: particularly applications concerning the structural health monitoring of bridge structures.
			\item Flood risk management schemes involving the introduction or removal of control structures such as leaky barriers .
		\end{itemize}
		However, when using the solver, as for all solvers and numerical schemes, care must be taken to ensure that the underlying assumptions are not violated in order to ensure sufficient modelling accuracy is achieved. In terms of further solver development, adding the capability to model vertical exchanges of momentum and non-hydrostatic pressure distributions would help to overcome some of the solvers current deficiencies. Although, addition of this further capability would also likely reduce the compatibility of the solver with existing numerical schemes. 
		
		\section*{Declaration of Competing Interest}
		The authors declare that they have no known competing financial interests or personal relationships that could have appeared to influence the work reported in this paper.
		
		\section*{Acknowledgements}
		This research was funded by the Engineering and Physical Sciences Research Council, United Kingdom grant number EP/T517914/1. Thank you also to Shannon Leakey for assisting by proof reading this document and providing valuable feedback. 
		
		}
		\bibliography{Test}
\end{document}